\documentclass[a4paper]{article}
\usepackage{makecell}

\usepackage{INTERSPEECH2022}

\title{EXPLORING REPRESENTATION LEARNING FOR SMALL-FOOTPRINT KEYWORD SPOTTING}
\name{Fan Cui, Liyong Guo, Quandong Wang, Peng Gao, Yujun Wang} 
\address{
  Xiaomi, Beijing, China
  }
\email{cuifan@xiaomi.com}

\begin{document}

\maketitle
\begin{abstract}
In this paper, we investigate representation learning for low-resource keyword spotting (KWS).
The main challenges of KWS are limited labeled data and limited available device resources. To address those challenges, we explore representation learning for KWS by self-supervised contrastive learning and self-training with pretrained model. First, local-global contrastive siamese networks (LGCSiam) are designed to learn similar utterance-level representations for similar audio samplers by proposed local-global contrastive loss without requiring ground-truth. Second, a self-supervised pretrained Wav2Vec 2.0 model is applied as a constraint module (WVC) to force the KWS model to learn frame-level acoustic representations. By the LGCSiam and WVC modules, the proposed small-footprint KWS model can be pretrained with unlabeled data. Experiments on speech commands dataset show that the self-training WVC module and the self-supervised LGCSiam module significantly improve accuracy, especially in the case of training on a small labeled dataset. 

\end{abstract}
\noindent\textbf{Index Terms}: keyword spotting, representation learning, self-supervised learning, self-training

\section{Introduction}
Keyword spotting (KWS) aims to identify the target keyword in audio segments, mainly used as the activation of voice assistants.
With the development of deep learning based speech technology, it is possible to deploy high-performance small-footprint keyword spotting system in commercial products (e.g., “Hey Siri” \cite{sigtia2018efficient},
“Alexa” \cite{sun2017compressed}, and “Okay Google” \cite{chen2014small}).

Recent keyword spotting methods focus on improving performance and reducing computation complexity to deploy on low-resource devices \cite{panchapagesan2016multi,bai2019time,zhuang2016unrestricted,yang2020multi,zhang2017hello,tucker2016model}. Especially, convolutional neural networks (CNN) based KWS methods show remarkable accuracy and efficiency \cite{sigtia2018efficient,sun2017compressed,yang2020multi,coucke2019efficient}. 2D CNN-based methods need very deep model to capture the dependency of different frequency bins and different frames. TC-ResNet \cite{choi2019temporal} is a fast and accurate KWS model, which applies temporal convolution network. However, those methods require large volumes of labeled data, which is time-consuming and expensive. Additionally, data distribution shifting between different recording environments leads to model mismatch. 

Self-supervised pretrained models have been investigated and shown significant improvements in different machine learning topics. BERT \cite{devlin2018bert} is a pretrained model that serves as a foundation for improving the accuracy of machine learning in natural language processing. Self-supervised speech recognition has been proposed with many approaches \cite{baevski2020wav2vec, van2018representation,schneider2019wav2vec}, which provide pretrained models that can be fine-tuned on downstream tasks for both
high-resource and low-resource conditions. Those self-supervised pretrained models are trained on large amounts of unlabeled data and can be used in two ways. First, they work as feature extractors and fine-tuned on a small amount of labeled data \cite{baevski2020wav2vec}. Second, they serve as teacher models to generate pseudo-label for training student models \cite{hubert,huang2018knowledge}. Depending on the downstream tasks,
the student model can be smaller. When the labeled data is enough to train the student model, pseudo-label can remove low-quality ground truth and avoid the student model learning implausible data. Additionally, pseudo-label can be generated for unlabeled data when the initial labeled data is not large enough to train a reliable model. In \cite{mazumder2021few}, pretrained embedding representation learning is proposed for few-shot KWS. However, the pretrained model is too large to deploy on low-resource devices.

In this study, we aim at exploring representation learning to improve the
small-footprint KWS model from two aspects. First, self-supervised pretrained Wav2Vec 2.0 \cite{schneider2019wav2vec} acoustic model is applied for teacher-student training, which uses unlabeled data to learn meaningful acoustic embedding representations. Instead of fine-tuning the Wav2Vec 2.0 for the downstream KWS model, we use it to extract hidden representations and force the encoder of the KWS model to learn similar speech representations.
Second, we design a self-supervised contrastive siamese networks (LGCSiam) for the proposed KWS model to learn similar utterance-level representations for similar audio samplers. The siamese networks are built upon recent progresses of contrastive learning in computer vision \cite{chen2020simple,chen2021exploring}. 
Siamese networks are weight-sharing neural networks with similar but different inputs. In this task, the random augments of audio sampler are used as the inputs of siamese networks.

Our contributions are as follows:
 \begin{itemize}
 	\item We propose a local-global contrastive siamese structure for KWS model to learning utterance representation from unlabeled data. An undesired result to siamese networks is all the outputs “collapsing” to a constant \cite{chen2021exploring}. To prevent from collapsing, the proposed structure generates positive pairs and negative pairs for contrastive learning. 
	\item We investigate teacher-student learning with self-supervised pretrained Wav2Vec 2.0 model for KWS model to learn acoustic representation.
	\item A light-weight KWS model TCANet is proposed, which uses temporal convolution networks in the encoder part, and adopt multi-head self-attention network as the decoder part.
	\item An evaluation of the representation learning for low-resource KWS.
\end{itemize}

The remainder of this paper is organized as follows: Section 2 introduces our proposed low-resource keyword spotting method. The experimental setup is presented in Section 3. The experimental results and analyses are shown in Section 4. Section 5 is the conclusion part.

\section{Methodology}
\begin{figure*}
  \centering
  \includegraphics[width=\linewidth]{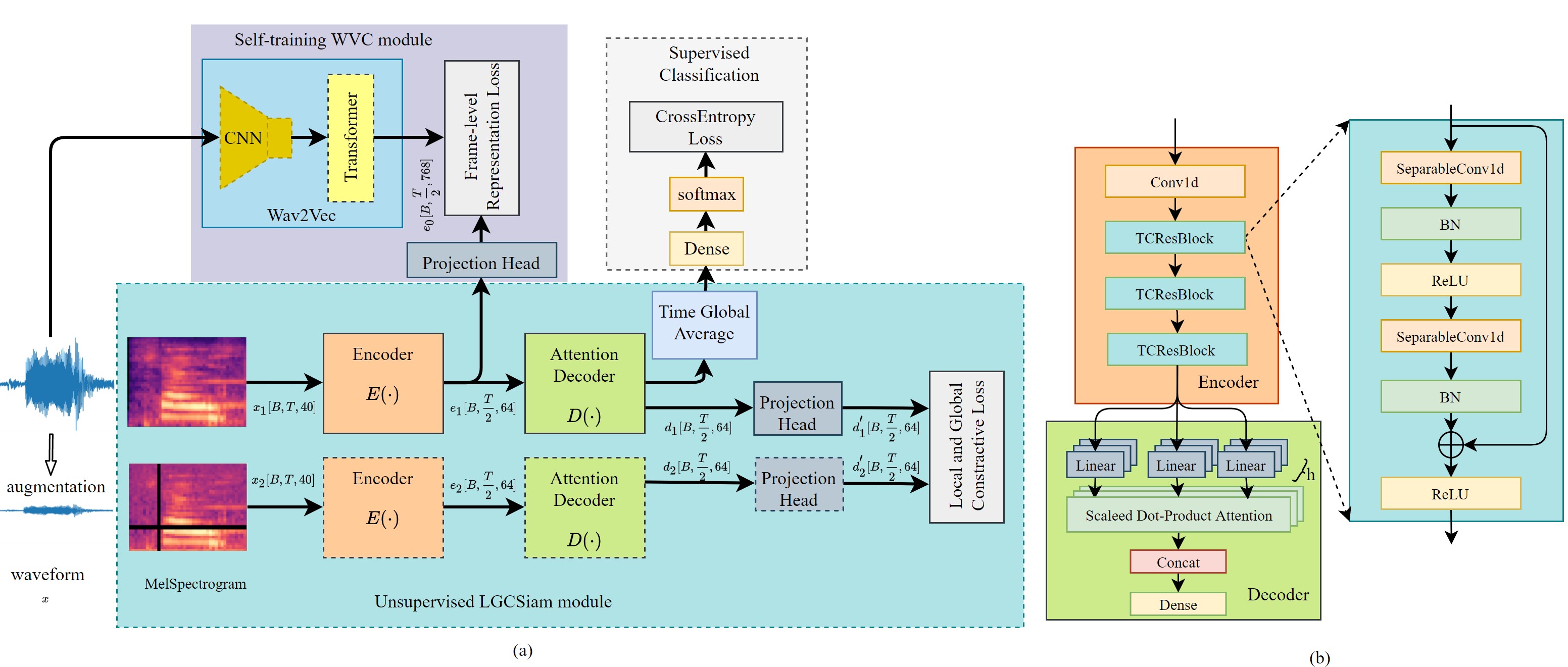}
  \caption{The proposed architecture. (a) the training workflow contains three parts: self-training WVC module, unsupervised LGCSiam module and supervised classification module. The WVC and LGCSiam modules are used for representation learning. The two branches of unsupervised LGCSiam module share the same weights. (b) shows the detail of TCANet model structure.}
  \label{fig:model1}
\end{figure*}

\begin{figure}[t]
  \centering
  \includegraphics[width=\linewidth]{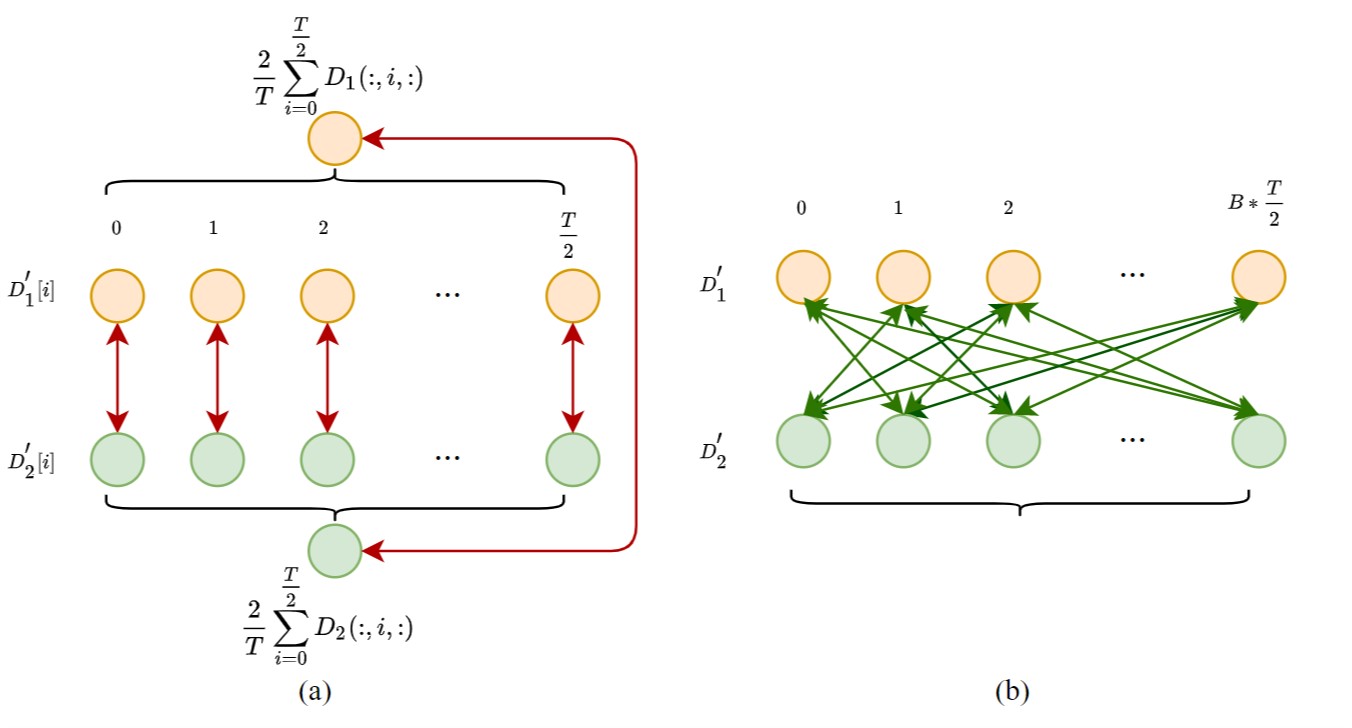}
  \caption{Positive pairs and negative pairs for local-global contrastive loss. (a) shows the global and local positive pairs of two corresponding audio samplers, (b) shows the negative pairs generated by different samplers or different timestamps.}
  \label{fig:model2}
\end{figure}

In this section, we present our system summarized in Figure~\ref{fig:model1}, contains self-training WVC module, self-supervised LGCSiam module and the light-weight TCANet model architecture.

\subsection{TCANet Model Architecture}
As shown in Figure~\ref{fig:model1}, the input waveforms are transformed to log-melspectrogram $x\in\mathbb{R}^{B\times{T}\times{40}}$, where ${B}$ represents batch size, ${T}$ is the maximum frame index, and the number of mel filterbanks is 40. The proposed TCANet model mainly contains the encoder $E(\cdot)$ and the decoder $D(\cdot)$. The encoder structure is similar to TC-ResNet, containing seven convolutional layers. The kernel size is 3 for the first layer and that of the other layers are 9. To reduce the complexity, the stride of the first layer is 2, which reduces time domain from $T$ to $\frac{T}{2}$. The output channel of each conv layer is 64. Rather than using normal 10.21437/Interspeech.2022-10558conv1d layer, we use separable conv1d to reduce parameters except the first layer. Compared with normal conv1d, separable conv1d can reduce parameter size from $9*64*64$ to $9*64*1+1*64*64$. Each of the convolutional layers followed by batch normalization and ReLU. After encoder, input $x\in\mathbb{R}^{B\times{T}\times{40}}$ is transformed to $e\in\mathbb{R}^{B\times{T/2}\times{64}}$.

The output embeddings of the encoder are fed into the decoder, which is a multi-head attention block \cite{vaswani2017attention}. The query, key and value are obtained by linear projections: $Q=e\cdot{W_Q}$, $K=e\cdot{W_K}$,  $V=e\cdot{W_V}$, where $W_Q,W_K,W_V\in\mathbb{R}^{64\times64}$. For each attention head, the output is calculated as $head_i=Softmax(\frac{Q_iK_i^T}{h_n})V_i$, where $h_n$ is $\frac{64}{num\_heads}$. The result of the decoder part is:
\begin{equation}
	D(e)=Concat(head_1,... head_n)W_o
	\label{eq1}
\end{equation}
The output of each attention head are concatenated, and projected by matrix $W_o\in\mathbb{R}^{64\times64}$. For supervised training, a global average pooling layer is applied to merge time domain to one frame. Finally, dense and softmax layer are adopted to get classification probability $p\in\mathbb{R}^{B\times{C}}$, $C$ is the number of class. The ground-truth is $y\in\mathbb{R}^{B\times{C}}$, where $y_{ij}=\{0,1\}$.
The cross entropy (CE) loss is adopted to optimize the
classification model:
\begin{equation}
	Loss_{CE}=-\sum_{i=1}^B\sum_{j=1}^Cy_{ij}log(p_{ij})
	\label{eq2}
\end{equation}

\subsection{WVC Module}

The Wav2Vec 2.0 model is a transformer-based neural network, which is pretrained on a large volume of unlabeled data. It is effective for semi-supervised downstream tasks training with small labeled dataset. 
It mainly contains four elements: feature encoder, context network, quantization module and pretraining contrastive loss \cite{baevski2020wav2vec}.

In the paper, we use Wav2Vec 2.0 base model\footnote{https://dl.fbaipublicfiles.com/fairseq/wav2vec/wav2vec\_small.pt}, which processes the latent feature vectors obtained by feature encoder through 12 transformer blocks in context network to increase the dimension from 512 to 768.
In our WVC module, input waveforms are fed into Wav2Vec 2.0 base model to get a context representation $e_0\in\mathbb{R}^{B\times{T/2}\times{768}}$, where $T$ is the maximum frame index with frame-shift 10ms. The KWS encoder is forced to learn frame-level acoustic representation, its output
$e_1\in\mathbb{R}^{B\times{T/2}\times{64}}$ should be similar to the
output of Wav2Vec 2.0 model ${e_0}$. Multi-layer perceptron (MLP) block is adopted as a projection head to transform $e_1$ to ${e_1}^{'}$:
\begin{equation}
	e_1^{'}=W_2\sigma(W_1\cdot{e_1})
	\label{eq3}
\end{equation}
Where $W_1\in\mathbb{R}^{64\times128}, W_2\in\mathbb{R}^{128\times768}$, $\sigma$ is a ReLU non-linearity. MSE loss is applied to measure the similarity of $e_1^{'}$ and $e_0$:
\begin{equation}
	Loss_{WVC}=\left\|e_1^{'}-e_0\right\|_2
	\label{eq4}
\end{equation}

\subsection{LGCSiam Module}
Two randomly augmented waveforms are transformed to log-melspectrogram $x_1\in\mathbb{R}^{B\times{T}\times{40}}$, $x_2\in\mathbb{R}^{B\times{T}\times{40}}$ token as the inputs of LGCSiam module. As shown in Figure~\ref{fig:model1}, the two branches of LGCSiam module share same weights in encoder and decoder parts. The two augmented samplers are processed by encoder $E(\cdot)$ and decoder $D(\cdot)$ networks to get representations $d_1\in\mathbb{R}^{B\times{\frac{T}{2}}\times{64}}$,$d_2\in\mathbb{R}^{B\times{\frac{T}{2}}\times{64}}$:
\begin{equation}
	d_1=D(E(x_1)), d_2=D(E(x_2))
	\label{eq5}
\end{equation}
A MLP projection head transforms the decoder representations as:
\begin{equation}
	d^{'}_1=W_4\sigma(W_3\cdot{d_1}), d^{'}_2=W_4\sigma(W_3\cdot{d_2})
	\label{eq6}
\end{equation}
Where $W_3\in\mathbb{R}^{64\times128}, W_4\in\mathbb{R}^{128\times128}$, $\sigma$ is a ReLU.

Our proposed Local-Global contrastive loss can force the KWS model to learning similar representations $d^{'}_1, d^{'}_2$. For calculating contrastive loss, we prepare positive and negative pairs, shown in Figure~\ref{fig:model2}. From local aspect, the corresponding frame in decoder results $d^{'}_1, d^{'}_2$ should servers as positive pairs $[{d^{'}_1(0,0,:),d^{'}_2(0,0,:)}]$, $[{d^{'}_1(0,1,:),d^{'}_2(0,1,:)}]$...$[{d^{'}_1(B,T,:),d^{'}_2(B,T,:)}]$. From a global aspect, the time-domain averaging of  $d^{'}_1, d^{'}_2$ should be a positive pair $[\frac{2}{T}\sum\limits_{i=0}^{\frac{T}{2}}{d^{'}_1(:,i,:)}, \frac{2}{T}\sum\limits_{i=0}^{\frac{T}{2}}{d^{'}_2(:,i,:)}]$. To avoiding the network learning useless same representations for all the inputs, the negative pairs should be designed. In this paper, the negative pairs are built by the decoder results of different timestamps $[{d^{'}_1(j_1,i_1,:),d^{'}_2(j_2,i_2,:)}]$, where ${i_1}\neq{i_2}$ or ${j_1}\neq{j_2}$. The global consistency loss is calculated as:
\begin{equation}
	L_{global}=\frac{1}{B}\sum_{j=1}^B\{ [\frac{2}{T}\sum\limits_{i=0}^{\frac{T}{2}}{d^{'}_1(j,i,:)} - \frac{2}{T}\sum\limits_{i=0}^{\frac{T}{2}}{d^{'}_2(j,i,:)}]^2\}
	\label{eq7}
\end{equation}
The local contrastive loss is calculated as:
\begin{equation}
L_{local}=-log\frac{exp(sim(d^{'}_1(i_1,j_1),d^{'}_2(i_1,j_1))/\tau)}{\sum_{i_2=0,j_2=0}^{i_2=B,j_2=\frac{T}{2}}exp(sim(d^{'}_1(i_1,j_1),d^{'}_2(i_2,j_2))/\tau)}
	\label{eq8}
\end{equation}
where ${i_1}\neq{i_2} || {j_1}\neq{j_2}$, temperature parameter  $\tau$ is 0.5, function $sim$ is cosine similarity. The local-global contrastive loss is:
\begin{equation}
    Loss_{LGCSiam}=L_{global}+L_{local}
	\label{eq9}
\end{equation}

The pretraining stage contains two steps. First, the encoder with WVC is optimized by $Loss_{WVC}$ with unlabeled data. Second, the encoder and decoder modules are trained with unlabeled data by loss function:
\begin{equation}
    L=\lambda_1Loss{LGCSiam} + \lambda_2Loss{WVC}
	\label{eq10}
\end{equation}
In the supervised fine-tuning stage, the classification module is added, the CE loss defined in equation~\ref{eq2} is combined with contrastive loss. The total loss function is:
\begin{equation}
    L=\gamma_1Loss{CE}+\gamma_2Loss{LGCSiam} + \gamma_3Loss{WVC}
	\label{eq11}
\end{equation}
where $\lambda_1$, $\lambda_2$, $\gamma_1$, $\gamma_2$, $\gamma_3$ are weights of the corresponding losses in different stages.


\section{Experimental setup}
\begin{figure}[t]
  \centering
  \includegraphics[width=\linewidth]{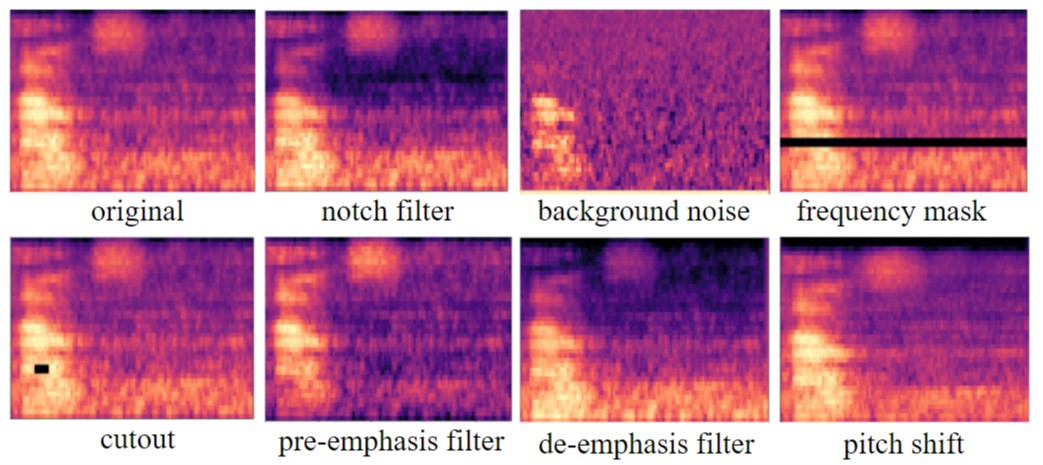}
  \caption{Different data augmentation operations.}
  \label{fig:data1}
\end{figure}

\subsection{Dataset}
The proposed models are trained and evaluated with speech commands dataset \cite{warden2018speech}, which contains 65000 one-second long audio utterances. Following the implementation of \cite{warden2018speech},  we split the dataset into training,
validation, and test sets.  Speech commands dataset contains 12 classes: “yes”, “no”, “up”, “down”, “left”, “right”, “on”, “off”, “stop”, “go”, unknown, or silence. We randomly select and clip background noise to generate background noise. To compare the representation learning on different domain datasets, unlabeled AISHELL dataset \cite{bu2017aishell} is used to pretrain the WVC module and LGCSiam module.

\subsection{Implementation Details}
All the models are trained and evaluated using Pytorch \cite{paszke2019pytorch} toolkit with a mini-batch of 128 samples. We use the stochastic gradient descent optimizer with momentum of 0.9. To avoid overfitting, we use a weight decay of 0.0001. The initial learning rate is set to be 0.1 and decayed by a factor of 3 when the validation accuracy does
not increase for 3 epochs. For equation~\ref{eq10}, $\lambda_1$, $\lambda_2$ are 0.1, 0.9. For equation~\ref{eq11}, $\gamma_1$, $\gamma_2$, $\gamma_3$ are 0.9, 0.05, 0.05.

\subsection{Data Augmentation}
Figure~\ref{fig:data1} shows the data augmentation operations randomly applied in our experiments:
\begin{itemize}
	\item Pre-emphasize: pre-emphasize an audio signal with a first-order autoregressive filter, $(y[n]=y[n]-coef*y[n-1]), coef\in\{0.95 \sim 0.99\}$
	\item De-emphasize: de-emphasize an audio signal with the inverse operation of Pre-emphasize.
	\item Pitch shift: shift the pitch of a waveform by $n_{steps}\in\{-5\sim5\}$ steps.
	\item Notch filter \& Peak filter: suppress or enhance the signal at a random frequency range.
	\item Adding background noise: the signal-to-noise ratio is between -5dB to 15dB.
	\item Frequency masking \cite{park2019specaugment}: the maximal number of masking frequency is 10. 
	\item Cutout: the maximal number of cutout frequency is 10 and the maximal number of time step is 10. 	
\end{itemize}

\subsection{Model setup}
\subsubsection{Baseline system}
\begin{itemize}
	\item TC-ResNet \cite{choi2019temporal}: TC-ResNet8 has one convolution layer and three residual blocks and {16, 24, 32, 48} channels for each layer. TC-ResNet14 expands the network by adding three residual blocks compared to TC-ResNet8. TC-ResNet8-1.5 and TC-ResNet14-1.5 are based on TC-ResNet8 and TC-ResNet14 with a channel multiplier of 1.5 expands the model to have {24, 36, 48, 72} number of channels respectively. 
	\item DS-CNN \cite{zhang2017hello}: DS-CNN
utilizes depthwise convolutional layer, which achieve high accuracy when memory and computation resources
are constrained.  DS-CNN-S, DS-CNN-M, and DS-CNN-L represent
small-, medium-, and large-size model, respectively. 
\end{itemize}
\subsubsection{Proposed system}
\begin{itemize}
\item TCANet: train the proposed TCANet model with labeled data. 
\item TCANet+WVC: pretrain TCANet by the WVC module with unlabeled data and fine-tuned with labeled data.
\item TCANet+LGCSiam: pretrain TCANet by the LGCSiam module with unlabeled data and fine-tuned with labeled data.
\item TCANet+WVC+LGCSiam: firstly, pretrain TCANet by the WVC module. Then, the LGCSiam module is trained with unlabeled data. Finally, fine-tune the whole model with labeled data.
\end{itemize}

\section{Results and Analysis}

\textbf{Impact of WVC and LGCSiam module pretrained with speech commands dataset}. In table~\ref{Semi-supervised_result}, we compare the original TCANet model trained with 100\% and 5\% speech commands dataset and the TCANet+LGCSiam, TCANet+WVC, TCANet+WVC+LGCSiam pretrained with unlabeled speech commands dataset and fine-tuned with 100\% and 5\% labeled data. Compared with supervised TCANet, pretrained models have a significant accuracy gain. The results show that the unsupervised pretraining helps the encoder and decoder to learning useful representation.

\textbf{Impact of WVC and LGCSiam module pretrained with AISHELL dataset}. In table~\ref{Semi-supervised_result_aishell}, we compare the original TCANet trained with 100\% and 5\% speech commands dataset and the TCANet+LGCSiam, TCANet+WVC, TCANet+WVC+LGCSiam pretrained with unlabeled AISHELL dataset and fine-tuned with 100\% and 5\% labeled speech commands dataset. From the results, we can see the models pretrained with AISHELL dataset also help to get a performance gain. The results show that the self-supervised pretraining on different domain dataset is useful for fine-tuning the downstream KWS task.

In table~\ref{final_result}, the proposed TCANet, TCANet+LGCSiam, TCANet+WVC,  TCANet+WVC+LGCSiam are compared  with the baseline systems. The results demonstrate the proposed system have a better performance with a smaller model size.

\begin{table}[!ht]
    \centering
    \caption{Comparing the influence of LGCSiam and WVC module for representation learning on Speech Commands (SC) dataset. We show the accuracies of TCANet+LGCSiam, TCANet+WVC and TCANet+WVC+LGCSiam pretrained with 100\% unlabeled SC dataset and fine-tuned with 100\% and 5\% labeled data .}
    \begin{tabular}{ l c c }
    \toprule
        \makecell[c]{pretrained on \\ 100\% unlabeled SC} & \makecell[c]{Acc. \\(100\% SC) } & \makecell[c]{Acc. \\(5\% SC) } \\
        \midrule
        TCANet(Without pretrain) & 96.85 & 88.24 \\
        TCANet+LGCSiam & 96.97 & 89.92 \\
        TCANet+WVC & 97.45 & 92.3 \\
        TCANet+WVC+LGCSiam & 97.5 & 92.52 \\
        \bottomrule
    \end{tabular}
    \label{Semi-supervised_result}
\end{table}

\begin{table}[!ht]
    \centering
    \caption{Comparing the influence of LGCSiam and WVC module for cross-domain representation learning. We show the accuracies of TCANet+LGCSiam, TCANet+WVC and TCANet+WVC+LGCSiam pretrained with 100\% unlabeled AISHELL dataset and fine-tuned with 100\% and 5\% labeled SC data.}
    \begin{tabular}{ l c c }
    \toprule
        \makecell[c]{pretrained on \\ 100\% unlabeled AISHELL} & \makecell[c]{Acc. \\(100\% SC) } & \makecell[c]{Acc. \\(5\% SC) } \\
                \midrule
        TCANet(without pretrain) & 96.85 & 88.24 \\
        TCANet+LGCSiam & 96.96 & 89.09 \\
        TCANet+WVC & 97.32 & 92.13 \\
        TCANet+WVC+LGCSiam & 97.45 & 92.33 \\
        \bottomrule
    \end{tabular}
    \label{Semi-supervised_result_aishell}
\end{table}

\begin{table}[!ht]
    \centering
    \caption{Comparison of various KWS models on the Speech Commands dataset.}
    \begin{tabular}{ l c c }
    \toprule
        Model & Acc. & Params \\
        \midrule
        DS-CNN-S & 94.4 & 24K\\
        DS-CNN-M & 94.9 & 140K \\        
        DS-CNN-L & 95.4 & 420K \\
        TC-ResNet8 & 96.1 & 66K\\
        TC-ResNet8-1.5 & 96.2& 145K \\
        TC-ResNet14 & 96.2 & 137K \\
        TC-ResNet14-1.5 & 96.6 & 305K \\
        TCANet & 96.85 & 65K \\
        TCANet+WVC+LGCSiam & 97.5 & 65K \\
    \bottomrule
    \end{tabular}
    \label{final_result}
\end{table}


\section{Conclusion}
In this paper, we investigate representation learning for keyword spotting with limited labeled data and limited device resources. We utilize self-training with pretrained acoustic model, and design a contrastive learning structure for keyword spotting. The experiments show the proposed representation learning modules are useful to improve accuracy with a small labeled dataset.

\clearpage
\bibliographystyle{IEEEtran}

\bibliography{mybib}


\end{document}